\documentclass[conference, 10pt]{IEEEtran}
\IEEEoverridecommandlockouts

\usepackage{cite}
\usepackage{amsmath,amssymb,amsfonts}
\usepackage{graphicx}
\usepackage{textcomp}
\usepackage{xcolor}
\usepackage[utf8]{inputenc}
\usepackage{amsmath}
\usepackage{algorithm}
\usepackage{algpseudocode}
\usepackage{setspace}
\usepackage{balance}

\definecolor{LightCyan}{rgb}{0.0,1,1}
\definecolor{LightGray}{rgb}{0.8,0.8,0.8}

\definecolor{blue}{rgb}{0.0,0.0,1}

\definecolor{red}{rgb}{1,0.0,0.0}

\definecolor{gray}{rgb}{0.3,0.3,0.3}

\def\BibTeX{{\rm B\kern-.05em{\sc i\kern-.025em b}\kern-.08em
    T\kern-.1667em\lower.7ex\hbox{E}\kern-.125emX}}

\begin{document}

\title{An Internet of Intelligent Things Framework for Decentralized Heterogeneous Platforms}

\author{\IEEEauthorblockN{Vadim Allayev}
\IEEEauthorblockA{\textit{Computer Science, CUNY--Queens College}\\
vadim.allayev94@qmail.cuny.edu}
\and
\IEEEauthorblockN{Mahbubur Rahman}
\IEEEauthorblockA{\textit{Computer Science, CUNY--Graduate Center \& Queens College}\\
mdmahbubur.rahman@qc.cuny.edu}
}


\maketitle
\thispagestyle{plain}

\begin{abstract}
Internet of Intelligent Things (IoIT), an emerging field, combines the utility of Internet of Things (IoT) devices with the innovation of embedded AI algorithms. However, it does not come without challenges, and struggles regarding available computing resources, energy supply, and storage limitations. In particular, many impediments to IoIT are linked to the energy-efficient deployment of machine learning (ML)/deep learning (DL) models in embedded devices. Research has been conducted to design energy-efficient IoIT platforms, but these papers often focus on centralized systems, in which some central entity processes all the data and coordinates actions. This can be problematic, e.g., serve as bottleneck or lead to security concerns. In a decentralized system, nodes/devices would self-organize and make their own decisions. Therefore, to address such issues, we propose a heterogeneous, decentralized sensing and monitoring IoIT peer-to-peer mesh network system model. Nodes in the network will coordinate towards several optimization goals: reliability, energy efficiency, and latency. The system employs federated learning to train nodes in a distributed manner, metaheuristics to optimize task allocation and routing paths, and multi-objective optimization to balance conflicting performance goals.
\end{abstract}

\begin{IEEEkeywords}
IoT, federated learning, metaheuristics, optimizations.
\end{IEEEkeywords}

\section{Introduction}\label{sec:intro}
Internet of Intelligent Things (IoIT), an emerging field, combines the utility of Internet of Things (IoT) devices with the innovation of embedded AI algorithms. It provides predictive and faster data analytics in IoT platforms thanks to machine learning algorithms which enable intelligent processing of huge amounts of sensor-generated data. However, it does not come without challenges, and struggles regarding available computing resources, energy supply, and storage limitations. In particular, many impediments to IoIT are linked to the energy-efficient deployment of machine learning (ML)/deep learning (DL) models in embedded devices. Naturally, running AI/ML on resource-constrained IoT nodes may pose such difficulties. Edge computing and embedded machine learning such as TinyML appear to be the hot topics to address these challenges. Looking further still, federated learning — or its variation split learning — may be the most optimal solution to approach this matter. With this, we may be able to reduce the communication burden on networks while enhancing scalability and promoting adaptability~\cite{Oliveira24}.

Research has been conducted to design energy-efficient IoIT platforms, but these papers often focus on centralized systems, in which some central entity processes all the data and coordinates actions. This can be problematic, e.g., serve as bottleneck or lead to security concerns. In a decentralized system, nodes/devices would self-organize and make their own decisions. Therefore, we propose to design a heterogeneous, decentralized sensing/monitoring IoIT mesh network. System performance is evaluated according to the criteria that it is 1) reliable, 2) energy efficient, and 3) has low latency. The system employs edge computing, federated learning, metaheuristic optimization, and multi-objective optimization. Federated learning allows nodes to learn from local data while training a shared model in a distributed manner, enabling privacy and scalability. Moreover, metaheuristic algorithms such as genetic algorithms and swarm intelligence can optimize resource/task allocation and routing/communication paths to minimize energy consumption. In particular, we focus on multi-objective optimization to balance our conflicting performance goals.

\subsection{Mission}
\textbf{Design a heterogeneous, decentralized peer-to-peer sensing and monitoring mesh network and optimize for reliability, energy efficiency, and latency.}

This paper explores:

\begin{enumerate}
    \item Descriptions and explanations of IoIT, edge computing, federated learning, metaherustic algorithms, mesh networks, and related concepts.
    \item Recently published papers pertaining to the above topics.
    \item The network architecture and implementation fitting for our proposed system.
    \item Multi-objective optimization, including what it is, why we need it, the objectives to optimize, variables to modify, and constraints to fulfill.
    \item Simulating multi-objective optimization for a decentralized mesh network and showcasing its results.
\end{enumerate}

\subsection{Use Cases}
This decentralized mesh network setup can be utilized to serve various use cases. For instance, in the more simple case, sensors could \textbf{detect local temperature or humidity}. Training the node models through federated learning would result in a holistic understanding of the temperature or humidity of that region. This would not necessarily require swarm intelligence or genetic algorithms, since the use case is simple, not resource intensive, and not time-critical. A more complex use case would be \textbf{identifying hot spots of cars or people in a city} in order to offer directions for the fastest routes and ultimately reduce traffic flow. This lends itself nicely to a mesh network, but would require constant updates of hot spot information throughout the city. In this case, swarm intelligence and genetic algorithms would be very applicable to optimize communication routes between nodes and reduce overhead. These are simply examples of how the system could be used. The main point is that they are both scenarios in which someone wants to monitor the information of some region with many nodes that act independently, and ensure all or most of that information propagates to and is accessible from every node.

Different aspects of the system can be modified to impact the reliability, energy efficiency, and latency of the system. This is what we explore in the multi-objective optimization and what we simulate for the evaluation. This optimization is particularly important since we are dealing with resource-constrained nodes, considering a heterogeneous system comprised of devices with different technical specifications/details, and catering to different circumstances. For instance, in one situation, latency may be held to a higher standard than the other two objectives. In another situation where the data exchanged is sensitive, reliability would be prioritized.

\section{Related Work}\label{sec:related}
\subsection{Related Algorithms}
In general, we have recently seen developments in energy-efficient algorithms for IoIT. For example, some researchers developed an energy-efficient and scalable routing technique for large distributed IoIT networks, specifically within a cloud-based Software-Defined Network (SDN) system. They utilized Genetic Algorithm (GA) and the swarm intelligence algorithms Particle Swarm Optimization (PSO) and Artificial Bee Colony (ABC). Their network scheduling technique applies Mobile Sink (MS) and optimizes the clustering of heterogeneous IoT nodes in the physical layer for MS to collect data. One concern would be the focus on a cloud-based Software-Defined Network, rather than edge computing~\cite{Udayaprasad24}. Another paper suggested metaheuristic-based routing, claiming that the generic energy-efficient routing framework in heterogeneous IoT is deficient. They propose two predictive models, one for energy-efficient node selection and another to address convergence issues. Though, these algorithms do seem more experimental and focus heavily on a metaheuristic approach~\cite{Rana21}. There have been other approaches too, for instance one paper that doesn't use GA, PSO, or ABC, but rather Proximal Policy Optimization (PPO)-based Deep Reinforcement Learning (DRL) to solve optimization problems to improve AI Generated Content (AIGC) quality and computation. The decentralized algorithm also integrates an LSTM (Long Short-Term Memory) model to improve its ability to handle temporal dependencies. Their methods take advantage of edge computing by offloading computing tasks to edge nodes, thereby optimizing task latency, energy efficiency, and load balancing. This could be a helpful paper to incorporate with respect to utilizing edge computing to the fullest, but would not coincide with optimizing the system for sensing/monitoring purposes~\cite{Zhang25}. The above papers focus more so on routing and scheduling, but could we apply and expand the concepts introduced in these papers to address monitoring as well?

\subsection{Federated Learning Applications}
To clarify, \textbf{federated learning (FL)} enables multiple devices to work together to train a machine learning model without sharing raw data. As FL has been showing promise, here are several papers that investigate this strategy to improve energy efficiency.

One paper presents a framework to minimize processing delay and reduce power consumption of Flying Ad-Hoc Networks (FANET) in Unmanned Aerial Vehicles (UAV) by exploiting Federated Reinforcement Learning. This is a study done based on an aerial environment, but we may be able to apply certain concepts into our own work. Nevertheless, the devices of the project may be homogeneous/standardized, which would be a source of conflict with our own goals~\cite{Grasso22}. Another promising paper outlines the development of a Decentralized Federated Learning (DFL) technique that performs distributed model training in Peer-to-Peer (P2P) manner by incorporating neighbor selection and gradient push. Their findings show a 57\% reduction in communication cost and 35\% in completion time. The algorithms would need to be adapted for our monitoring purposes and environment of sensors, but the technique itself seems enticing~\cite{Liao24}.

Something to note, in general FL does inherently require some level of coordination, since the local data from each sensor would need to be aggregated to a shared model, i.e., it would involve some central aggregator, like a server. However, there are fully-decentralized alternatives. Namely, P2P (peer-to-peer) FL (utilized in the previous paper mentioned) with gossip learning, and Blockchain-based FL.

\subsubsection{Decentralized FL Variations}
P2P FL with gossip learning serves to overcome the bottleneck problem that comes from a centralized system. Gossip learning is scalable, fault-tolerant, with minimized communication overhead, but may encounter issues with consistency, security, and propagation time~\cite{naik2023introduction}. One study observed gossip learning operating at a similar level to federated learning, despite its lack of a central controlling entity. Gossip learning is able to converge in a practically realistic time frame, and demonstrates its capacity to compete with typical FL. The authors believe gossip learning could be improved with more sophisticated peer sampling methods~\cite{hegedHus2021decentralized}.

Blockchain-based FL is another option, utilizing blockchain, a public, trusted and shared ledger running on a P2P network. One paper explains its improved security, but also its issues with communication cost and resource allocation. They suggest ways we may remedy these, for instance reward-based training~\cite{Nguyen21}. However, a different survey paper of blockchain-based FL posits that the approach is increasingly impractical and difficult, in large part due to the significant computational power required, tricky tradeoffs between performance and other factors like energy efficiency, and even potential issues with security as well~\cite{Qu22}.

For the sake of reliability, energy efficiency, and latency, P2P FL with gossip learning looks like the most compatible approach, as it is lightweight and straightforward.

\subsection{Related Infrastructure}
Regarding the system infrastructure, a multi-hop mesh network seems the most appropriate and practical, considering our aim to work on decentralized system with heterogeneous devices. Edge computing also seems more fitting to use as opposed to cloud, as it is currently favored in IoT and offers reduced latency and increased privacy~\cite{Oliveira24}. However, a cloud-based approach remains a consideration, as it would permit us to circumvent limited computational power and storage capacity constraints. We could even implement LoRa architecture, which has a cloud component. 

One paper that seems relevant to our goal discusses a heterogeneous Edge-IoT mesh network with Multi-Hop-Over-The-Air update technology that enables auto-configuration of devices and quick deployment of services. This distributed and collaborative ecosystem is a great demonstration of the infrastructure we seek to employ~\cite{Carnevale21}. Another paper explored a heterogeneous, semi-distributed algorithm for traffic demand forecasting using graph neural networks (GNNs) and leveraging data center computation via the cloud. They explain that decentralizing the entire GNN operation led to excessive node communication and overhead, so this solution prevents that and enhances scalability, though it is not fully decentralized~\cite{Nazzal24}. Lastly, another paper focused on decentralized edge computing for Community Mesh Networks (CMNs), using lightweight virtualization and Information-Centric Networking (ICN) to incorporate in-network caching, name based routing, and to develop smart heuristic. The focus of the paper is on service delivery rather than monitoring, but we could take inspiration from the heuristic approach (relating to the other metaheuristic paper) and edge-IoT approach in our own work~\cite{Lertsinsrubtavee18}.

\section{System Model}\label{sec:sysmodel}
\subsection{Assumptions} 
An IoT system inherently has resource constraints, including limited storage, computing power, and energy. Heterogeneous devices/sensors may be from different vendors and may have different sensor types, battery capacities, and power source differences. However, all must work with Zigbee technology and protocols. The environment may have interference, so some messages may fail to send. However, nodes within each other's transmission range are assumed to be able to communicate with each other.

\subsection{Background Information}
\textbf{Federated learning (FL)} enables multiple devices to work together to train a machine learning model without sharing raw data. \textbf{Peer-to-peer FL (P2P FL)} is a fully decentralized version of typical FL without a central server with which the nodes communicate. A \textbf{mesh network} refers to a decentralized network topology in which nodes can connect to other nodes directly to create a robust and fault-tolerant wireless coverage area. \textbf{Metaheuristic algorithms} are higher-level procedures that help find solutions to optimization problems. They accomplish this by finding sufficiently good solutions to the problem, which drastically reduces the time it takes to solve, as opposed to trying to find the exact most optimal solution, which may take considerably longer. \textbf{Swarm intelligence} refers to algorithms inspired by the collective behavior of decentralized, self-organized systems, for instance ant colonies or flocks of birds, for the sake of optimization. \textbf{Genetic algorithms} are similar, they mimic natural selection over many generations to find optimal or near-optimal solutions to a problem. \textbf{Edge computing} refers to focusing on processing and storing information locally, closer to the source of data, as opposed to externally on the cloud. Namely, operating at the level of sensors/nodes which reduces latency and bandwidth usage. \textbf{Zigbee} is a wireless communication protocol designed for low-power, low-bandwidth devices. It is especially prevalent in IoT applications. It is used in a mesh network environment, permitting devices to communicate with each other directly or indirectly through other devices in the network, thus creating a more robust and reliable system.

\subsection{Network Architecture}
When deliberating between edge and cloud computing, and considering current advances in technology, edge presents itself as the better approach. Since it does not need to access a cloud server externally, it inherently has less actuation time and faster decision-making capabilities. Moreover, we want to limit the involvement of a centralized element in this project.

The primary components of our system involve physical sensing nodes, Peer-to-Peer Federated Learning (P2P FL), and metaheuristic optimization algorithms. The system starts with sensor-equipped devices that are placed around a given region. These can be any type of sensor, from tracking temperature to traffic. These sensors would monitor and collect data from their environment and train their own model on that local data. Next, devices/nodes would discover each other via Zigbee and utilize P2P FL with gossip learning (which has proved itself to be effective~\cite{Liao24, naik2023introduction, hegedHus2021decentralized} to share their models with their neighbors. This way, no centralized entity is involved and nodes operate on their own, updating their model over time with P2P FL as data converges. For more complex use cases, we implement swarm intelligence algorithms in order to optimize peer discovery, routing paths for communication, and resource usage; primarily Ant Colony Optimization (ACO) but possibly also Particle Swarm Optimization (PSO). Furthermore, since this is a heterogeneous system, different devices may have different physical capabilities and may inherently prioritize different metrics (e.g., low latency vs. low energy consumption). Thus, we incorporate multi-objective optimization to improve performance and compatibility. These considerations may help to maximize reliability and minimize energy consumption and latency, enabling us to cater the system to the use case's requirements.

For the simulation, we will consider a use case where we want to have a good understanding of the temperature of some region, represented as an array of dimensions \(R \times R\), such that our region will be split up into \(R^2\) squares/subregions. We will generate a solution array for what we want the temperatures to look like. Since temperature in nearby subregions cannot realistically be extremely different, we will make sure adjacent subregions do not vary more than a few degrees. Each of the \(N\) nodes will "train its local model" by "detecting" the temperature in its subregion. The node will accurately monitor the temperature in its subregion, but will assume neighboring subregions have the same temperature. This accounts for the slight degree of error characteristic in these circumstances. When a node is aggregating model data it just received from a neighbor to its own model, it will find the average between each subregions. In general, for this simulation we focus more so on communication between nodes rather than sensing and processing data.

\section{Multi-Objective Optimization}\label{sec:multi-objective}

\subsection{Objectives}
Our aim to create a fully decentralized sensing/monitoring system inherently poses some interesting design challenges. Namely, without the presence of a central controlling unit, all the communicating and processing transpires on the level of the nodes. The system we propose is also heterogeneous, so each node may have different storage and memory specifications and may prioritize different metrics (e.g., low latency vs. low energy consumption). As we originally mentioned, we are focusing on (1) reliability, (2) energy efficiency, and (3) low latency. Due to the conflicting objectives present, we deemed it necessary to formulate a multi-objective optimization problem. For the following equations, \(n\) represents the number of nodes and \(N\) represents the set of nodes used in the simulation.

\subsubsection{Reliability}
Reliability represents the correct delivery of data. In terms of a simulation, we can think of it as the likelihood that a node completes its tasks successfully and yields accurate results. The tasks of a node include: collecting sensor data, exchanging information with peers, and aggregating neighbor model data to its own model. For the sake of simplicity, we may not focus on processing time at the node level, since the process of a node gathering data from the environment and updating its P2P FL model is largely out of our control and up to the protocols of the hardware device and library, respectively. Communicating data between nodes reliably, rather than processing data, is the priority.

Simply, we attribute the reliability to \textbf{the number of total number of successful unique message deliveries, \(SM\), with respect to the total number of unique messages sent, \(TM\)}. Considering unique messages specifically is important, since there is a chance that the message may drop and be resent, thereby making it successful, albeit on two or more attempts. In the equations below, \(R\) represents the reliability of the system, and \(r(i)\) represents the reliability of a generation.

\begin{equation}
    R = \frac{\sum_{i=1}^{g} r(i)}
    {g}, \text{ } r(i) = \frac{\sum_{j=1}^{n} SM_j/TM_j}
    {n} \leftarrow\text{in gen } i \label{eq:reliability}
\end{equation}



\subsubsection{Latency}
Latency generally refers to the delay before a data transfer, but in the context of a system with its own unique goals, its representation becomes more nuanced. One common representation is end-to-end latency, which broadly refers to the total time it takes for a signal to travel from its source to its destination across a network, including transmission and processing time. However, this is a decentralized system. Rather than observing the delays involved with every node communicating with a base station, we simply have nodes that communicate with each other for the sake of building more accurate models. This makes measuring latency more complicated. (Note, similarly to reliability, we may not incorporate node data processing time in the calculation, but rather only the delays associated with transmission/propagation.)

One approach is to measure how long it would take for every node to receive model updates from the majority of all the other nodes. Simply, how long it would take for all the nodes to be trained on data originating from some specified proportion of the other nodes in the network.

Alternatively, we could measure how long it would take for the system to become "accurate", such that each node's local model is "close enough" to the actual data of the region that the system is trying to monitor. Moreover, to ensure that the system has reached a competent level of understanding of the region's data, we can also mandate that the average node accuracy must be above a certain threshold as well. Thus there would be a minimum and an average accuracy requirement.

Another option is to designate one or more nodes as endpoints, whose aggregated models data would theoretically be collected by users at those locations. In this case, the latency would represent how long before all of the specified endpoints have "accurate" data. However, this resembles end-to-end latency, characteristic of centralized systems, since these endpoints serve as the general destination.

To stick with the decentralized nature of the project and offer a holistic and simplistic approach, we consider latency as the time it takes for the system to become generally "accurate" (with respect to the region's actual data). Below, \(min\), \(avg\), and \(max\) represent the minimum, average, and maximum node accuracy of a generation (as percentages), respectively.

\begin{equation}
    L = g \text \quad \text{if } ( avg_g > \psi \text{ } \lor \text{ } avg_g+1\geq max_g)\text{ }\land \text{ } min_g > \theta
\end{equation}

where \(g =\) the number of simulated generations, \(\psi\) represents the required threshold for the average accuracy of the system, \(\theta\) represents the minimum requirement for the accuracy of all nodes in the system, and \(\theta \leq \psi < 1\). Notice that there are two possibilities for the average accuracy requirement. It either must be greater than \(\psi\), or within 1\% of the max accuracy. This contingency ensures that a solution may be found even in scenarios where nodes cannot detect the entire region, i.e., there are subregions that no node can reach. This is acceptable since there is also a \(min\) requirement, by which all nodes must have an accuracy of at least \(\theta\). To clarify, several messages may be sent in one generation, and a generation is a generic unit of time used for the simulation, e.g., one hour, one day.

This approach to measuring latency can be likened to the time it takes to reach model convergence. That is to say, how long before the models of the endpoint nodes more or less stabilize, and would only minimally benefit from further training. Therefore, we can think of latency in our system simply as \textbf{the number of generations until model convergence}.

\subsubsection{Energy Efficiency}
This objective tracks the energy consumed in relation to the amount of work being done. Simply, we look at \textbf{the energy expended in communicating data}. Although reliability and latency may become more or less of a priority, we should always strive to minimize energy consumption (thereby maximizing energy efficiency), considering the resource-constrained nature of IoT devices. Along with latency, energy efficiency can be optimized with the help of swarm intelligence or genetic algorithms, but for our calculation, we look generally at how much energy is consumed for a given simulation.

To calculate the energy consumption of a node carrying out a single task, we use \(E = v \times i \times t\), where \(E\) is the energy consumed (in J) over \(t\) seconds with voltage \(v\) (in V) and current \(i\) (in Amp). A node can transmit or receive. The energy consumed during these modes may be represented as \(Tx\) and \(Rx\), respectively. The energy consumed when idling is minimal and we consider it negligible. We measure the energy consumption of the system, \(EC\), as the average of \(e\), the energy expenditure of a node in each generation, with respect to the number of generations. Thus, the energy consumption is represented as: 

\begin{equation}
    EC = \frac {\sum_{i=1}^{g} e(i)} {g} \label{eq:energy-consumption}, \text{ } e(i) = \frac{\sum_{j=1}^{n} Tx_j + Rx_j}{n} \leftarrow\text{in gen } i
\end{equation}


\subsection{Decision Variables}
In the context of optimization problems, these are the inputs that we adjust to obtain various results. By changing one decision variable at a time and keeping the rest of the variables constant, we can observe the effects it brings about and ascertain the most optimal value(s) for each variable. Here are the variables that we modify:
\begin{enumerate}
    \item \textbf{Sharing Frequency}: number of neighbors to which each node may transmit messages (during each generation). Sharing with more nodes would expend more energy but reduce latency since it would lead to faster model convergence.
    \item \textbf{Resend Threshold}: the maximum number of messages that a node is willing to resend in a generation. Resending a message to the same node would increase reliability, but may lead to increased latency or energy consumption in the case that it takes multiple retries.
    \item \textbf{Communication Strategy}: the criteria that nodes use to decide which neighbors to share their models with. Among these, we can choose neighbors (a) randomly, sending to any neighbor within range, or (b) to the least interacted, prioritizing neighbors the node has communicated with the least.   
\end{enumerate}
We must also consider the \textbf{number of nodes} in the system, the \textbf{the size of the region} in square meters, and the \textbf{node placement} around the region, namely in a random or uniform distribution. Different quantities and dimensions may impact the results of the optimization, especially considering the clustering of nodes and the inherent limitations of transmission and detection range. Therefore, we also run different simulations (with different decision variable values) for different scenarios of node amounts, region sizes, and node placement.

\subsection{Constraints}
In the context of optimization problems, constraints are the minimum requirements that solutions to the system must satisfy. These are non-negotiable standards that terminate a simulation if violated. For our purposes, here are the constraints:
\begin{enumerate}
    \item \textbf{Activity}: All nodes in the system must remain active, i.e., none can reach 0\% battery.
    \item \textbf{Energy}: The energy of the system (i.e., the average energy of the nodes) cannot drop below \(\phi\), where \(50\%< \phi < 100\%\).
    \item \textbf{Connectivity}: All nodes must be connected to every other node in some way.
\end{enumerate}

\subsection{Pareto-Optimal Front}
The goal of the multi-objective optimization is to obtain the \textbf{Pareto front}, also known as the Pareto frontier. This is the set of all Pareto-efficient solutions. A \textbf{Pareto-efficient solution} is a solution where it is impossible to improve any objective without making at least one other objective worse. In our case, a solution represents a simulation with a particular setup of decision variables. So, a Pareto-efficient solution would be a solution where you could not improve its reliability, energy efficiency, or latency without decreasing something else. For example, some of these solutions may be optimized for reliability, and others for other objectives. Thus, we may end up with a versatile set of solutions that are all the best in their own way, consequently offering many options for system specifications.

\section{Implementation}\label{sec:impl}
For the evaluation of this paper and its associated results, we simulate its behavior in Python. 
The simulation utilizes the Numpy, NetworkX, Matplotlib, and SciPy libraries to create a network with nodes, simulate gossip learning over many generations to propagate model data between nodes, and repeat the process many times over to generate different results for different network setups. We also script several functions to visualize the data.

For the system, we consider battery-operated devices with microcontrollers. Therefore, devices cannot regenerate their battery by means of solar energy. Regarding hardware emulation, we would use relatively inexpensive physical sensing nodes capable of Zigbee communication (and thus operating on the 2.4GHz band). SimpleLink 32-bit Arm Cortex wireless MCUs would be sufficient for this purpose. Namely, CC1310~\cite{CC1310}, CC1352R~\cite{CC1352R}, and CC2652R~\cite{CC2652R}. Since this is a decentralized environment, there is no centralized entity. For the federated learning aspect, we would implement P2PFL, a decentralized federated learning library in Python~\cite{P2PFL}.
\section{Evaluation}\label{sec:eval}
\subsection{Experimental Setup}
To test the system, we simulated a network environment in Python. To conduct the multi-objective optimization, we set up an experiment based on the simple use case example. To reiterate, let us consider a scenario in which the president of a Homeowners Association is in charge of some neighborhood with dimensions \(R \times R\) (in meters) whose temperature he wants to measure holistically. Implementation-wise, this would look like a 2D matrix that represents the temperatures in the different locations (or "subregions") of the region, relative to its position in the matrix. In order to obtain a more accurate measure of the temperature in the region, he decides to install sensors across the neighborhood to obtain regional temperature data. He may decide to do so by installing them uniformly atop the street lamps of the neighborhood. He may instead implore residents to install their own sensors on their mailboxes, which would lead to a more random distribution of the sensors across the neighborhood, with the incentive that they could easily access an accurate representation of the regional temperature. In either case, this scenario lends itself nicely to our system model; a decentralized, heterogeneous system.

Before starting, we can make some assumptions:
\begin{itemize}
    \item Each sensor has knowledge of nearby sensors within its transmission range. We can assume that there is some initialization stage during which the sensors discover and take note of each other.
    \item Each node will have the following properties:
    \begin{itemize}
        \item 50m transmission range
        \item 30m detection range
        \item 1000J maximum energy
    \end{itemize}
    \item All nodes start at maximum energy.
    \item The three constraints mentioned before will be fulfilled (otherwise the simulation will terminate).
    \item The goal of the system is that all nodes have an accurate enough model (i.e., meeting the average and minimum system accuracy requirements), in order that each node has a good understanding of the general temperature of the region.
\end{itemize}

This involves creating a Node and P2PNetwork class in Python. Our determiner for success will be a simulation's performance with respect to the three objectives we discussed earlier. A simulation will be structured as follows:
\begin{enumerate}
    \item Generate the values for \(area\_data\), a 2D matrix of integers that represents the temperatures of the region.
    \item Create the nodes of the system, either with uniform or random placement around the region.
    \item Build and store a NetworkX graph that will keep track of all the neighbors of a node (i.e., the nodes within its transmission range). 
    \item Calculate the connectivity of the graph. If not sufficiently connected as outlined by the Connectivity Constraint, redo steps 2-4.
    \item Generate local data. This represents the nodes monitoring the temperature in their local subregions and storing that data internally.
    \item All nodes exchange model data with neighboring nodes, in accordance with its \(sharing\_frequency\), \(resend\_threshold\), and \(communication\_strategy\). The duration of this process is considered 1 round or "generation," which can be likened to some measurement of time in the real world (e.g. 1 hour). This step represents the P2P FL with gossip learning and develops the model of each node.
    \item Repeat step 6 until model convergence is reached or until \(max\_rounds\) have passed.
\end{enumerate}

As mentioned before, we can determine different results by keeping all decision variables constant except for one which we change. To accomplish this, we create several nested for loops, with the variable of each loop modifying and pertaining to a different decision variable. Specifically, 
\begin{itemize}
    \item Sharing frequency will be tested at values 1 through 5 neighbors/generation,
    \item Resend threshold at values 0, 5, 10, ..., 50 messages/generation, and
    \item Communication strategy at either random or least-interacted.
\end{itemize}
After each simulation, we save the results so that we could visualize them all together. For redundancy and robustness, there will be one more layer of a for loop, which will cause every simulation setup to run multiple times.

To investigate the effects of different area sizes, amounts of nodes, and node distributions, we conduct this experiment of running a plethora of decision-variable-changing simulations for the following scenarios:
\begin{itemize}
    \item \(300m^2\) \& 81 nodes (uniform distribution)
    \item \(300m^2\) \& 100 nodes (random distribution)
    \item \(500m^2\) \& 121 nodes (uniform distribution)
    \item \(500m^2\) \& 250 nodes (random distribution)
\end{itemize}

\subsection{Pseudocode}
We implement the following algorithm (Algorithm (\ref{algo:gossip})) to carry out the gossip learning stage of the simulation, which constitutes what happens during 1 generation of the simulation. Notice the use of the communication strategy and sharing frequency decision variables:

\begin{algorithm}[h]
\small
\caption{Gossip Learning in Decentralized P2P Sensing Networks}
\begin{algorithmic}[1]
\Procedure{GossipLearning}{$communicationStrategy$, $sharingFrequency$}
    \State $round \gets round + 1$
    \State Initialize $reliabilityData[1..n\_nodes] \gets [0, 0, ..., 0]$
    \State Initialize $energyData[1..n\_nodes] \gets [0, 0, ..., 0]$
    
    \State Shuffle nodes randomly to simulate random behavior
    
    \For{each $node$ in $nodes$}
        \If{$communicationStrategy = \text{"least-interacted"}$}
            \If{$|node.neighbors| \leq sharingFrequency$}
                \State $selected \gets node.neighbors$
            \Else
                \State $selected \gets node.neighbors[0..sharingFrequency-1]$
            \EndIf
            \State $node.neighbors \gets node.neighbors[sharingFrequency..] + node.neighbors[0..sharingFrequency-1]$ \Comment{Rotate neighbors}
        \Else \Comment{Random selection strategy}
            \State $selected$ $\gets$ $\text{Random sample of }$ $min(sharingFrequency, |node.neighbors|)$
            $\text{neighbors from } node.neighbors$
        \EndIf
        
        \State $successfullySent \gets 0$
        \State $totalSent \gets 0$
        
        \For{each $neighborId$ in $selected$} \Comment{Send model and save results}
            \State $target \gets nodes[neighborId]$
            \State $(numSuccessful, totalMessages$, $selfEnergyConsumed,$ 
            $neighborEnergyConsumed) \gets node.sendModel(target)$
            \State \Comment{Keep track of reliability and energy consumed}
            \State $successfullySent \gets successfullySent + numSuccessful$
            \State $totalSent \gets totalSent + totalMessages$
            \State $energyData[node.nodeId] \gets energyData[node.nodeId] + selfEnergyConsumed$
            \State $energyData[neighborId] \gets energyData[neighborId] + neighborEnergyConsumed$
        \EndFor
        
        \State $reliabilityData[node.nodeId] \gets successfullySent / totalSent$
    \EndFor
    
    \State \Return $reliabilityData, energyData$ 
\EndProcedure
\end{algorithmic}
\label{algo:gossip}
\end{algorithm}

Additionally, see below what occurs during the process of a node sending its model data to a neighboring node. Note, Algorithm~\ref{algo:send_model} below is used by Algorithm 1 above. Also, this procedure occurs at the node level. The process of sending model data showcases:
\begin{enumerate}
    \item two main transmission modes (with and without acknowledgments), 
    \item probabilistic success rates for message transmission,
    \item retry logic for failed transmissions, which makes use of the resend threshold decision variable,
    \item realistic energy consumption calculations (impacted by node hardware specifications) for both the sender and the receiver,
    \item and the Activity Constraint, which is triggered if a node becomes inactive.
\end{enumerate}

\begin{algorithm}
\scriptsize  
\caption{Send Model Between Nodes in P2P Sensing Network}
\begin{algorithmic}[1]
\Procedure{SendModel}{$self, target$}
    \State Define $ActivityConstraint()$ that exits simulation if a node becomes inactive
    \State $numPackets \gets |self.sensedLocations|$
    \State $numMessages \gets self.calculateNumMessages(numPackets)$
    \If{$self.resendThreshold = 0$} \Comment{No acknowledgments needed}
        \State $successful[1..numMessages] \gets$ Random values where $P(success) = self.messageSendSuccessRate$
        \State $numSuccessful \gets \sum{successful}$
        \State $numUnsuccessful \gets numMessages - numSuccessful$
        \State $time \gets self.messageSendTime(numMessages)$
        \State $selfEnergyConsumed \gets$ $self.voltage \times$ $self.txCurrent \times time$
        \State $self.consumeEnergy(selfEnergyConsumed)$
        \If{$\neg self.active$}
            \State $ActivityConstraint()$
        \EndIf
        \State $receiveTime \gets self.messageSendTime(numMessages - numUnsuccessful)$
        \State $targetEnergyConsumed \gets target.voltage \times target.rxCurrent \times receiveTime$
        \State $target.consumeEnergy(targetEnergyConsumed)$
        \If{$\neg target.active$}
            \State $ActivityConstraint()$
        \EndIf
    \Else \Comment{Acknowledgments required}
        \State $sentSuccessful[1..numMessages] \gets$ Random values where $P(success) = self.messageSendSuccessRate$
        \State $ackSuccessful[i] \gets$ Random value where $P(success) = self.ackSendSuccessRate$ if $sentSuccessful[i]$, else $False$
        \State $initialSent \gets \sum{sentSuccessful}$
        \State $initialAcks \gets \sum{ackSuccessful}$
        \State $numUnsuccessful \gets numMessages - initialAcks$
        \State $retries \gets 0$
        \State $additionalAcksSent \gets 0$
        \State $additionalAcksReceived \gets 0$
        \While{$numUnsuccessful > 0$ \textbf{and} $retries < self.resendThreshold$}
            \State $retries \gets retries + 1$
            \If{$Random() \leq self.messageSendSuccessRate$} \Comment{Retry message goes through}
                \State $additionalAcksSent \gets additionalAcksSent + 1$
                \If{$Random() \leq self.ackSendSuccessRate$} \Comment{ACK message goes through}
                    \State $additionalAcksReceived \gets additionalAcksReceived + 1$
                    \State $numUnsuccessful \gets numUnsuccessful - 1$
                    \State Update first $False$ in $ackSuccessful$ to $True$
                \EndIf
            \EndIf
        \EndWhile
        \State \Comment{Calculate energy consumption}
        \State $selfTransmitEnergy \gets self.voltage \times self.txCurrent \times self.messageSendTime(numMessages + retries)$
        \State $selfReceiveEnergy \gets self.voltage \times self.rxCurrent \times self.messageSendTime(initialAcks + additionalAcksReceived, 5)$
        \State $selfEnergyConsumed \gets selfTransmitEnergy + selfReceiveEnergy$
        \State $self.consumeEnergy(selfEnergyConsumed)$
        \If{$\neg self.active$}
            \State $ActivityConstraint()$
        \EndIf
        \State $targetReceiveEnergy \gets target.voltage \times target.rxCurrent \times self.messageSendTime(initialSent + additionalAcksSent)$
        \State $targetTransmitEnergy \gets target.voltage \times target.txCurrent \times self.messageSendTime(initialAcks + additionalAcksSent, 5)$
        \State $targetEnergyConsumed \gets targetReceiveEnergy + targetTransmitEnergy$
        \State $target.consumeEnergy(targetEnergyConsumed)$
        \If{$\neg target.active$}
            \State $ActivityConstraint()$
        \EndIf
        \State $successful \gets ackSuccessful$
        \State $numSuccessful \gets \sum{ackSuccessful}$
    \EndIf
    \State $target.updateModel(self, successful)$
    \State \Return $numSuccessful$, $numMessages$, $selfEnergyConsumed$, $targetEnergyConsumed$
\EndProcedure
\end{algorithmic}
\label{algo:send_model}
\end{algorithm}

\subsection{Trial Run}
Before showing the results of the above four scenarios, here is a trial run simulation to showcase the process and outcomes of a single simulation (as opposed to hundreds meant to optimize our objectives). The first trial run is for a \(300m \times 300m\) region with 100 nodes that are randomly distributed. Here is what that looks like Figure~\ref{fig:trial-graph}:
\begin{figure}
    \centering
    \includegraphics[width=.8\linewidth]{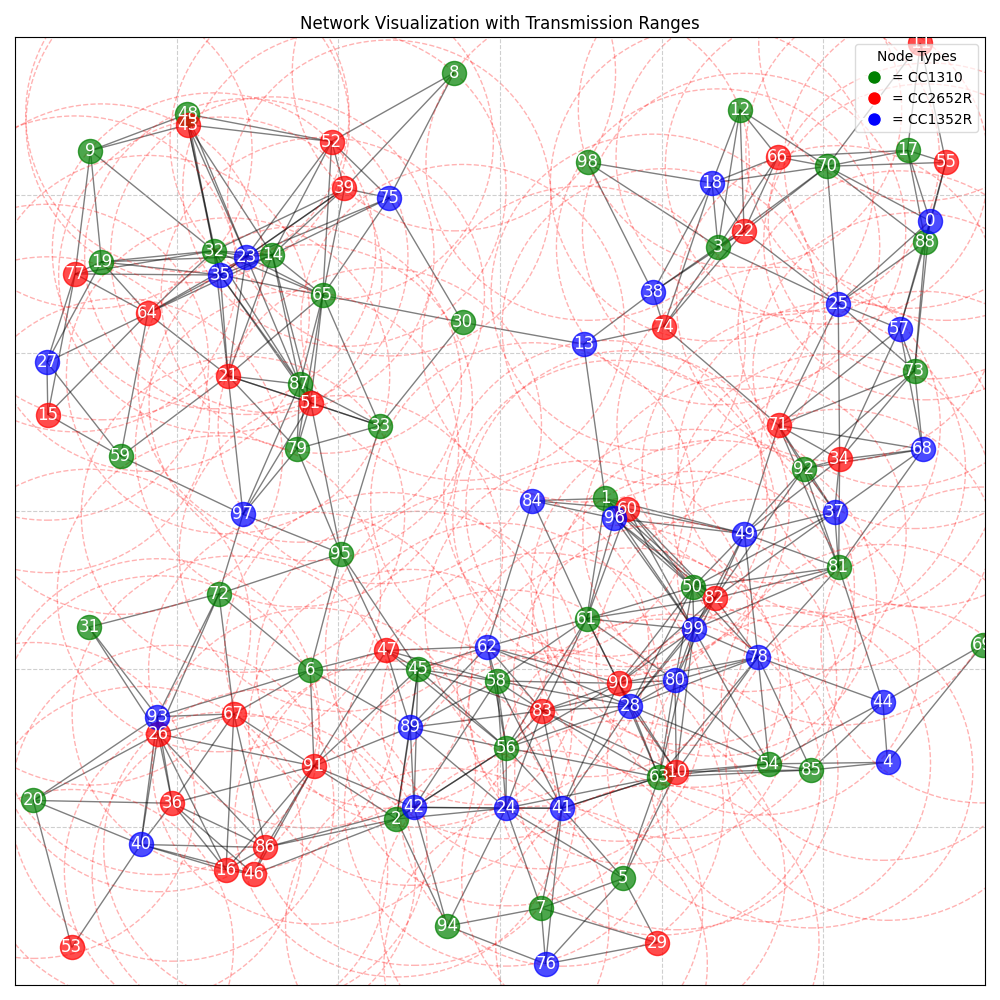}
    \caption{Trial Run Graph with Random Nodes}
    \label{fig:trial-graph}
\end{figure}

Notice the different coloring of nodes, which represent different MCUs, highlighting the system's heterogeneity. Notice also that each node is connected. The dotted red circles show the transmission ranges of each node.

\begin{figure}[H]
    \centering
    \includegraphics[width=.8\linewidth]{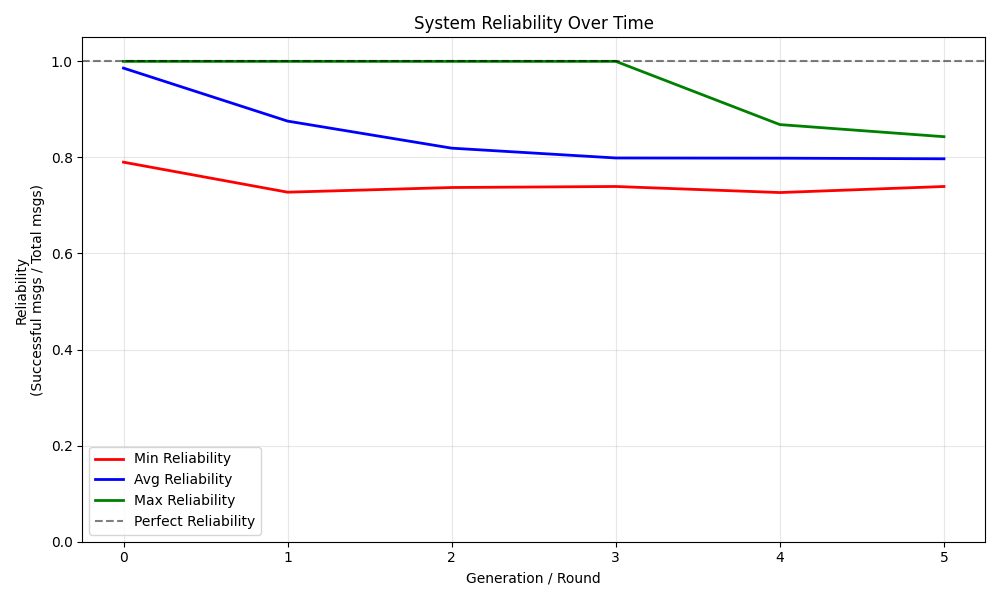}
    \caption{Trial Run Reliability}
    \label{fig:trial-reliability}
\end{figure}

Now, for the decision variables, let us set the sharing frequency to 3 neighbors/generation, the resend threshold to 5 messages/generation, and the communication strategy to least-interacted. The results of the simulation are as follows, Figure~\ref{fig:trial-reliability}

This is the reliability per generation. The average and maximum reliability understandably dip over time. In the beginning, when models are undeveloped and data is small, there are only a few messages that need to be transmitted to convey all of a node's data. However, over time, models develop, and nodes need to send more information, leaving more room for potential error. The resend threshold is only 5 messages per generation, which is evidently insufficient to maintain a perfect or near perfect reliability. 

\begin{figure}
    \centering
    \includegraphics[width=0.8\linewidth]{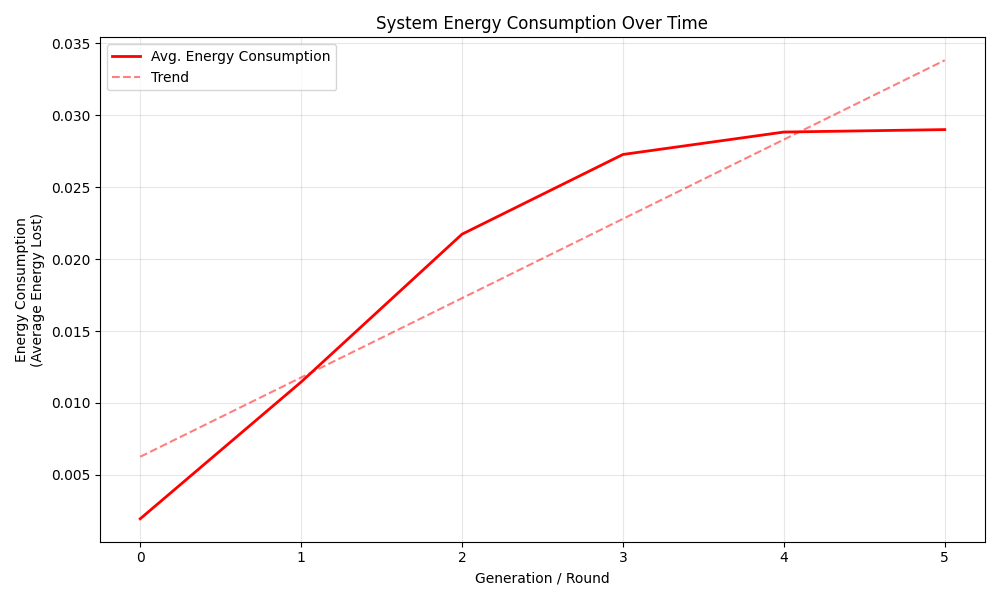}
    \caption{Trial Run Energy Efficiency}
    \label{fig:trial-energy}
\end{figure}
Figure~\ref{fig:trial-energy}: Energy efficiency naturally increases over the generations for the same reason; more data, and therefore more messages, needs to be sent in later generations. We see the energy consumed level out towards the end, since by that point the nodes have very developed models, without many improvements.

\begin{figure}
    \centering
    \includegraphics[width=.8\linewidth]{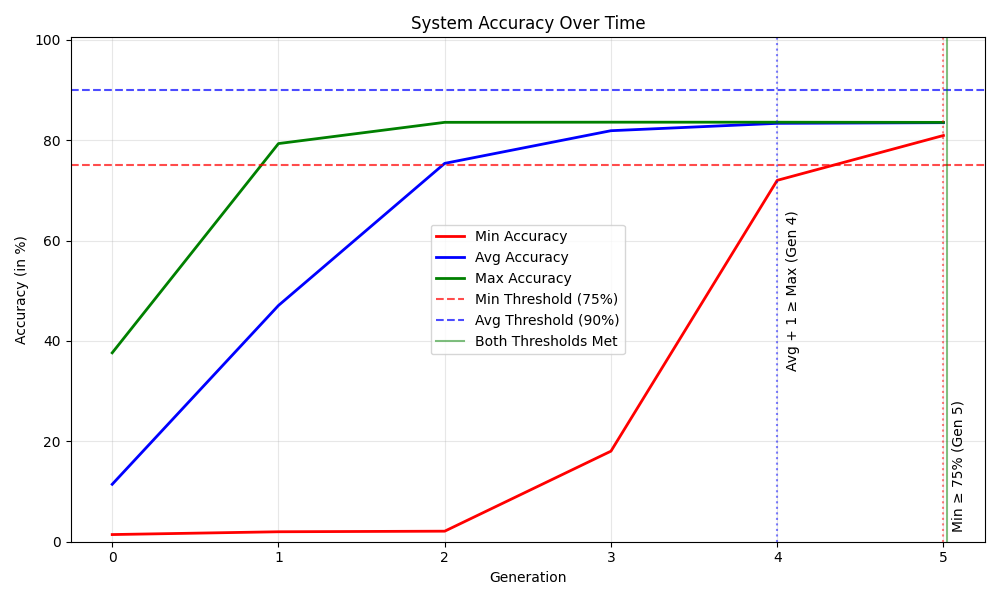}
    \caption{Trial Run Latency}
    \label{fig:trial-latency}
\end{figure}
Figure~\ref{fig:trial-latency}: We can observe the average accuracy requirement being fulfilled in generation 4, and the minimum accuracy requirement in generation 5. Moreover, the average and maximum accuracy seem to progress quickly, but the minimum accuracy lags behind. This could indicate that a focus on communicating with nodes with less developed models may be beneficial.

\subsection{Results}
\subsection{81 Uniform Nodes}
\begin{figure}
    \centering
    \includegraphics[width=.8\linewidth]{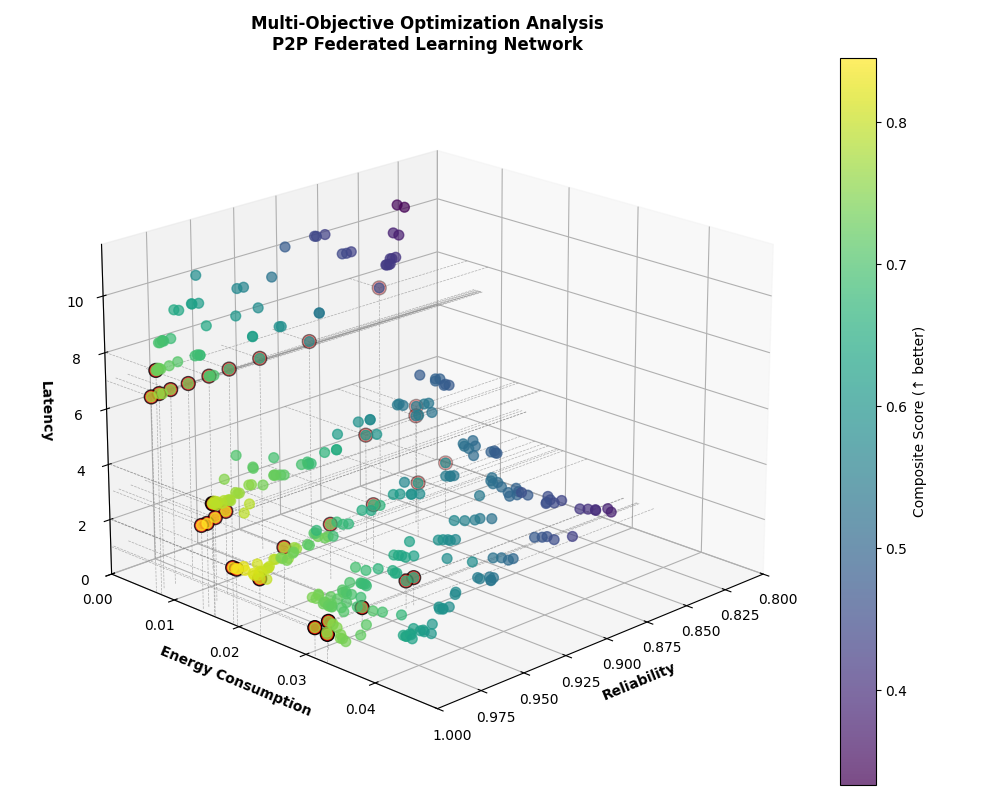}
    \caption{\(300m^2\) Area \& 81 Uniform Nodes Results}
    \label{fig:uniform300m81n1}
\end{figure}

For each experiment, we plot the results of the simulations on a 3D graph, with the objectives as the axes. Here is the first experiment, with area size \(300m^2\), 81 nodes, and uniform node distribution: Figure~\ref{fig:uniform300m81n1}.

The points with a red outline represent the Pareto-efficient solutions. As you can see, there are a variety of effective results, the consequence of different simulation setups. The color of each point is an overall "performance score," simply calculated by normalizing the objective results and combining them.

\begin{figure}
    \centering
    \includegraphics[width=.8\linewidth]{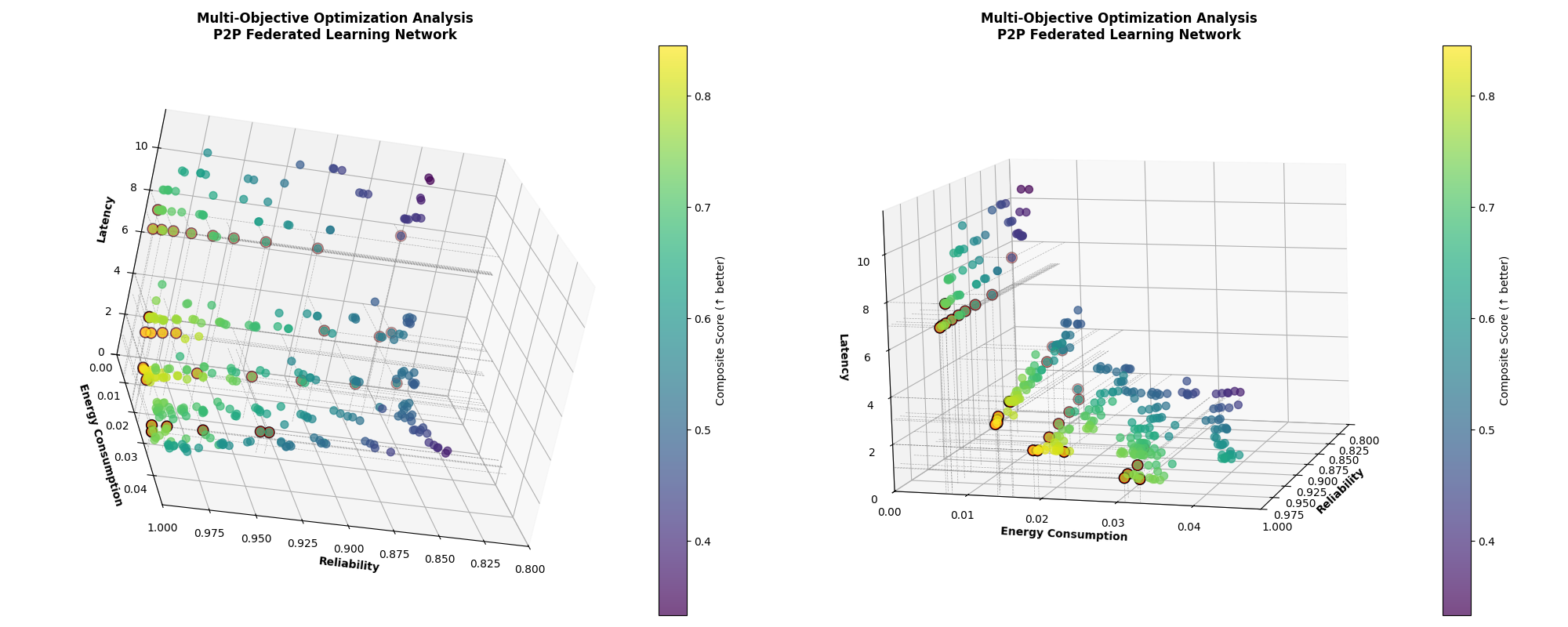}
    \caption{\(300m^2\) Area \& 81 Uniform Nodes Results Other Angles}
    \label{fig:uniform300m81n2and3}
\end{figure}

Figure~\ref{fig:uniform300m81n2and3} shows the same experiment from different angles. Notice that even darker color points, which are technically a worse performance score, are still part of the Pareto frontier. Even if they are not among the best performing setups, they are the best option for what they optimize. 

\begin{figure}
    \centering
    \includegraphics[width=0.8\linewidth]{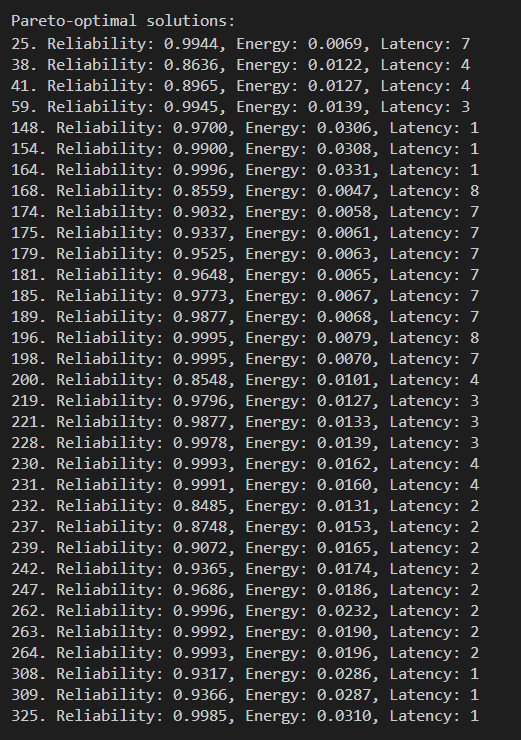}
    \caption{\(300m^2\) Area \& 81 Uniform Nodes Pareto-Efficient Solutions}
    \label{fig:uniform300m81nSolutions}
\end{figure}

Figure~\ref{fig:uniform300m81nSolutions} showcases the Pareto-efficient solutions from this experiment, listing each solution's performance in each objective. To reiterate, the reliability represents the average proportion of successful messages sent per generation, energy consumption represents the average energy consumed by a node per generation, and latency represents the generation in which the system converges, i.e., the average and minimum system accuracy requirements are fulfilled. We can observe the differences between simulations. For example, simulation \#168 has a relatively low reliability of 85.59\% and high latency with the simulation lasting for 8 generations, but very little energy consumption. On the other hand, simulation \#325 had a near perfect reliability at 99.85\% and needed only 1 generation to converge, but expended more than 6 times the amount of energy. Different system setups would be used for different scenarios.

\begin{figure}
    \centering
    \includegraphics[width=0.95\linewidth]{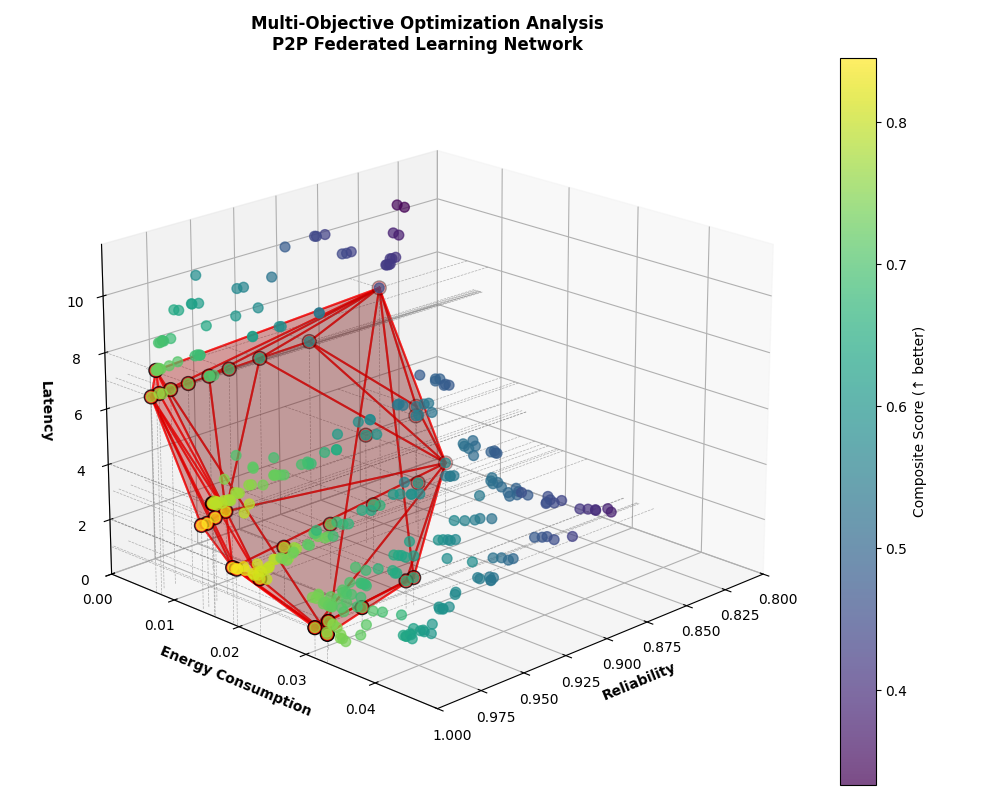}
    \caption{\(300m^2\) Area \& 81 Uniform Nodes Results with Mesh}
    \label{fig:uniform300m81nMESH1}
\end{figure}

\begin{figure}
    \centering
    \includegraphics[width=0.95\linewidth]{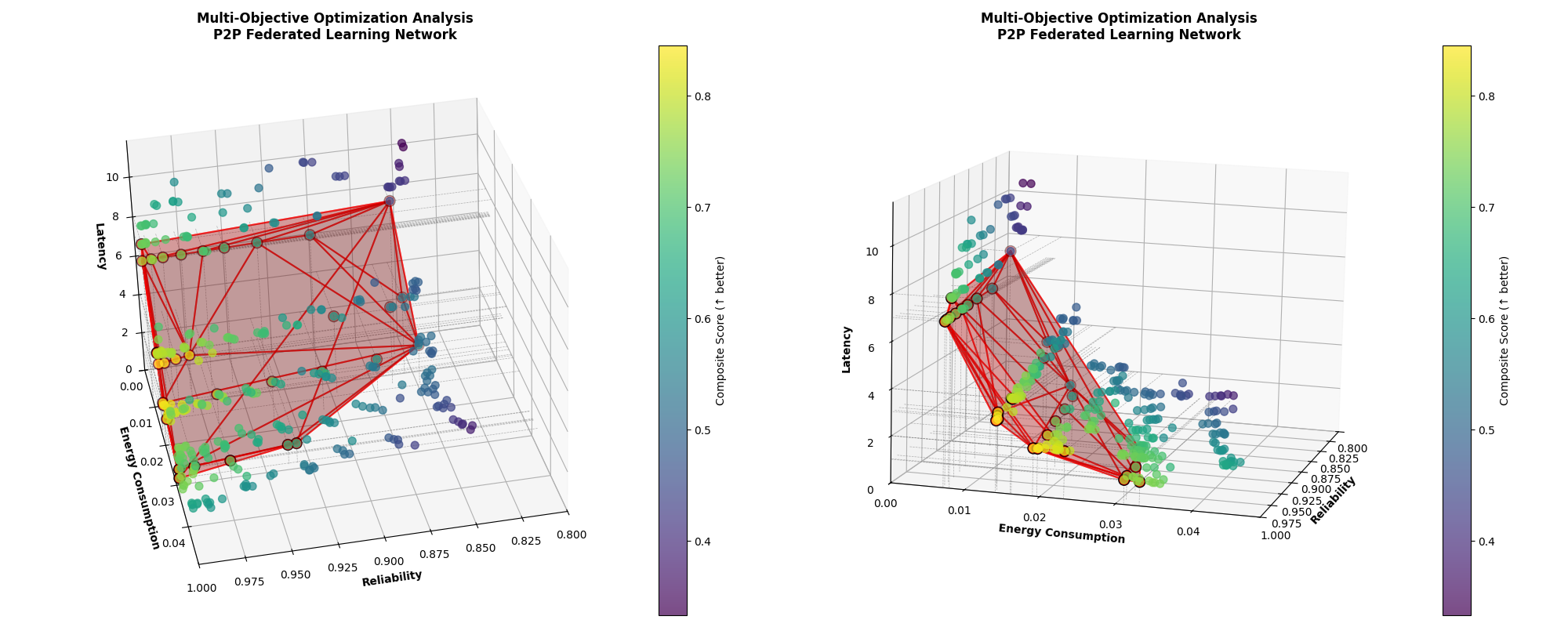}
    \caption{\(300m^2\) Area \& 81 Uniform Nodes Results with Mesh Other Angles}
    \label{fig:uniform300m81nMESH2and3}
\end{figure}

In Figures~\ref{fig:uniform300m81nMESH1} and~\ref{fig:uniform300m81nMESH2and3}, you can see a visualization of the Pareto frontier, overlayed by the points of the graph. Any location on the mask is theoretically and realistically possible to emulate, enabling individuals to obtain even more personalized reliability, energy efficiency, and latency results.

\subsection{250 Random Nodes}
\begin{figure}
    \centering
    \includegraphics[width=0.8\linewidth]{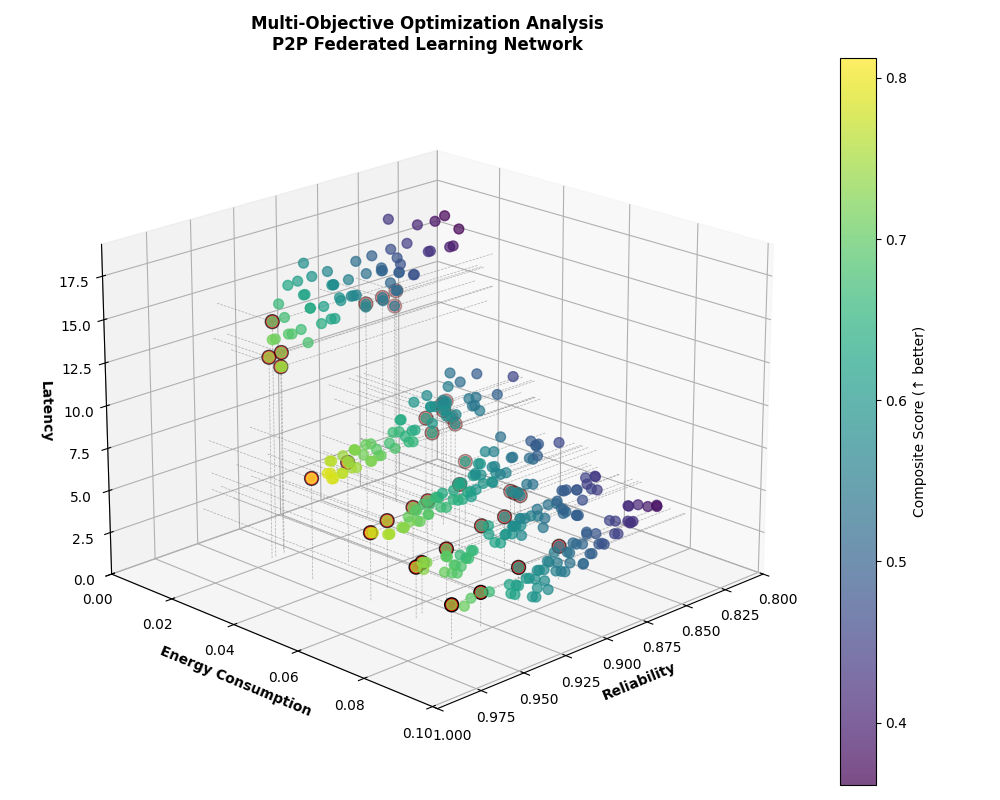}
    \caption{\(500m^2\) Area \& 250 Random Nodes Results}
    \label{fig:random500m250n1}
\end{figure}

\begin{figure}
    \centering
    \includegraphics[width=0.8\linewidth]{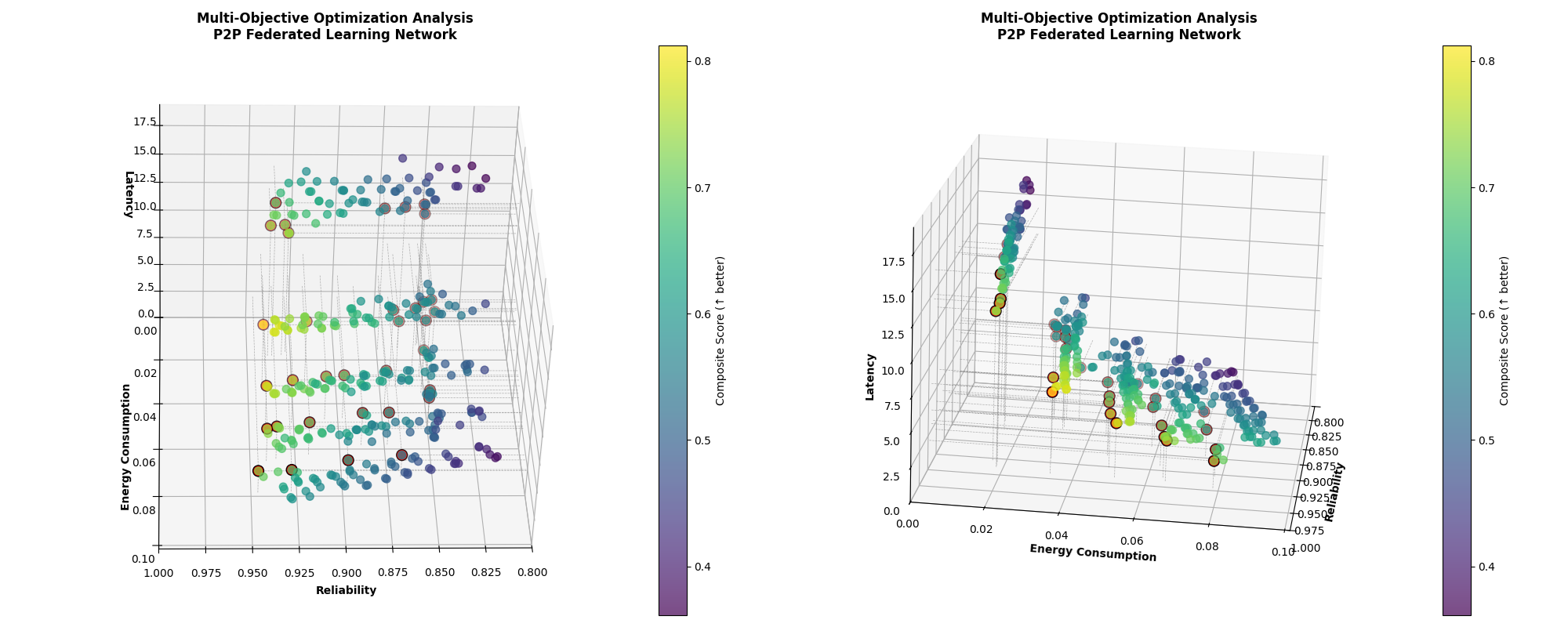}
    \caption{\(500m^2\) Area \& 250 Random Nodes Results Other Angles}
    \label{fig:random500m250n2and3}
\end{figure}

To investigate a larger-scale problem, here is the last experiment, with area size \(500m^2\), 250 nodes, and random distribution: Figures~\ref{fig:random500m250n1} and~\ref{fig:random500m250n2and3}

The distribution of points on the graph are relatively similar, but the reliability across the board has decreased, and energy consumption has just about doubled. Some simulations did have really low latency, but the upper limit increased a lot. We can also see that there are more scattered clusters of points. 

\begin{figure}
    \centering
    \includegraphics[width=0.95\linewidth]{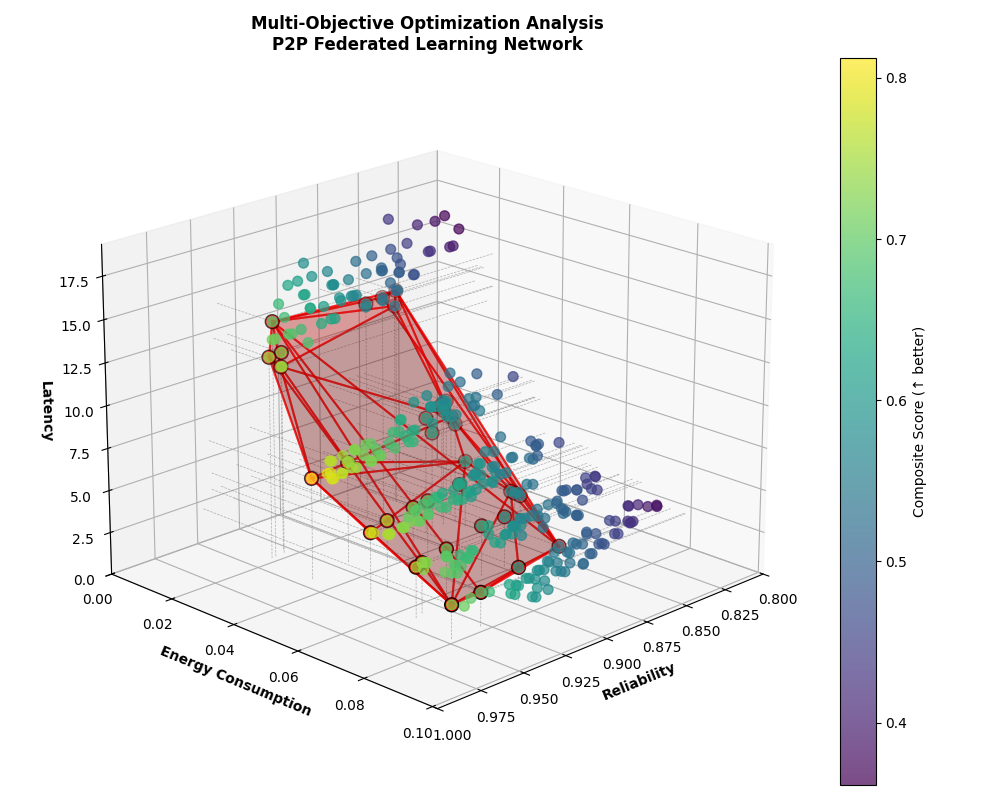}
    \caption{\(500m^2\) Area \& 250 Random Nodes Results with Mesh}
    \label{fig:random500m250nMESH1}
\end{figure}

\begin{figure}
    \centering
    \includegraphics[width=0.95\linewidth]{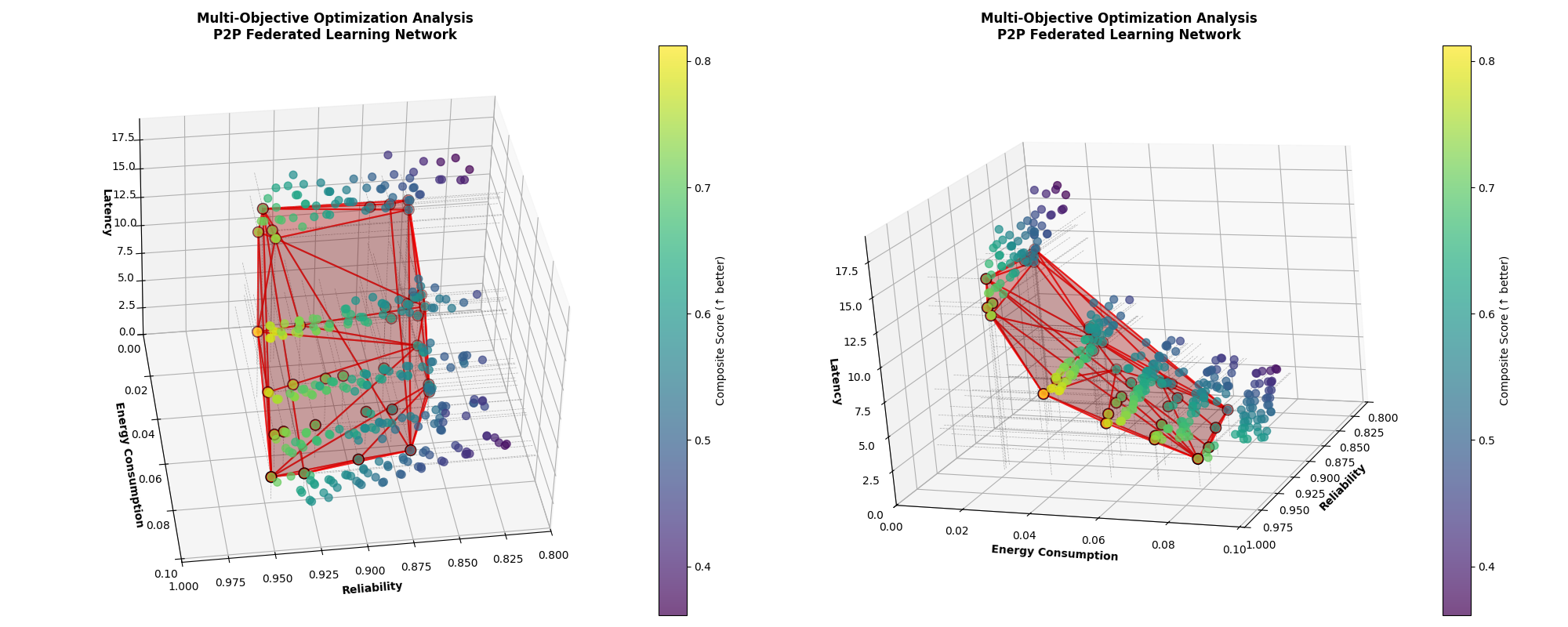}
    \caption{\(500m^2\) Area \& 250 Random Nodes Results with Mesh Other Angles}
    \label{fig:random500m250nMESH2and3}
\end{figure}

Figures~\ref{fig:random500m250nMESH1} and~\ref{fig:random500m250nMESH2and3}: Generally, uniform node placement throughout the region led to decreased latency and energy consumption. We also needed significantly fewer nodes to fulfill the Connectivity Constraint and convergence if we placed them uniformly. A uniform distribution meant that each node was equidistant from one another and therefore had several neighbors within its range guaranteed. However, there was still always a case where the system could converge in 1 generation, regardless of the area size or node distribution. Moreover, the reliability was overall about the same.

\section{Conclusion}
\subsection{Takeaways}
IoIT is an emerging new technology with significant applications. Made possible thanks to edge computing, federated learning, and TinyML (embedded ML), we are approaching a time of innovative advancements in health, fitness, security, agriculture, manufacturing, transportation, and more. The current challenges with this framework include limited computing resources, energy supply, and storage limitations, the consequence of running AI/ML on inherently resource-constrained and often battery-powered IoT devices.

Metaheuristics is a crucial paradigm in the context of optimization, enabling us to solve complex NP hard problems sufficiently and in much less time. Among them are genetic algorithms inspired by natural selection and swarm intelligence which emulates collective behavior of natural systems such as ants colonies, bird flocks, and bee swarms.

Multi-objective optimization entails optimizing different objectives in order to find the Pareto-efficient solutions. By modifying the decision variables and enforcing the constraints, we were able to simulate our heterogeneous, decentralized peer-to-peer mesh network in different ways to yield different results. This system model is one that could certainly be utilized with the prospect of IoIT, as it addresses the issues with centralized systems; namely, the potential for the centralized entity serving as a bottleneck or being taken down, rendering the entire system inoperable. 

\subsection{Future Works}
There are many ways we could improve the simulation. One method is to implement more varied and nuanced decision variables. For instance, regarding the communication strategy decision variable (which determines the neighbors a node chooses to talk to), we could also have the deciding factor be neighbors with the most energy, or neighbors who have the least updated models. This would offer even more versatile results, especially since before deciding who to send to, nodes would need to send an additional smaller message to neighboring nodes, requesting their energy level or model development, respectively. This would lead to more potentially dropped packets and slightly more energy expended, but may decrease latency as system models may converge faster.

Another idea is to refine the simulation by considering more aspects of the system. We could consider Euclidean distances between nodes, so messages have a higher drop rate over longer distances. We could consider a "trust" score to combat malicious actors, so nodes could decide for themselves that another node is suspicious (e.g., sending false information, sending too many requests) and reroute accordingly. We could even consider varying node transmission ranges, though that may reveal difficulties like the hidden terminal problem.

Implementing metaheurstic algorithms would also be a great step. For example, some environments may have rivers or mountains or inaccessible areas, making node distribution less straightforward. We could use a genetic algorithm to ascertain the most optimal node placement in the system. Or, during the actual simulations, to determine the best routing paths and improve fault tolerance, we could use swarm intelligence like Ant Colony Optimization to dynamically coordinate node communication.

Additionally, we could incorporate techniques from the papers in the Related Works section. For example, we could leverage the decentralized Proximal Policy Optimization-based Deep Reinforcement Learning approach outlined in one paper to offload computational tasks to edge servers \cite{Zhang25}, or incorporate neighbor selection and gradient push, which reduced communication cost and completion time in a different paper that worked on their own decentralized FL technique \cite{Liao24}.

Of course, the best course of action would be to purchase the physical hardware and run these simulations in an outdoor environment, with the actual P2P FL software, and see how the system holds up. Through my simulation, the system model looks practical. But considering this as a practical real system necessitates real-world testing.

\bibliographystyle{IEEEtran}
\balance
\bibliography{IEEEabrv,mybib}

\end{document}